\documentclass[aps,pra,twocolumn,groupedaddress,showpacs]{revtex4-1}
\usepackage{amssymb,amsmath}
\usepackage[varg]{txfonts}
\usepackage{graphicx}
\usepackage[colorlinks,linkcolor=blue,citecolor=blue,urlcolor=blue]{hyperref}

\renewcommand{\vec}[1]{\mathbf{#1}}
\mathchardef\mhyphen="2D
\newcommand{\dd}{\mathrm{d}}
\newcommand{\iu}{\mathrm{i}}

\begin{document}

\title{Disentangling multipole contributions to collective excitations in fullerenes}
\author{M. Sch{\"u}ler}\email[]{michael.schueler@physik.uni-halle.de}
 \author{J. Berakdar} \author{Y. Pavlyukh}  \affiliation{Institut
  f\"{u}r Physik, Martin-Luther-Universit\"{a}t Halle-Wittenberg,
  06099 Halle, Germany}

\date{\today}

\begin{abstract}
  Angular resolved electron energy-loss spectroscopy (EELS) gives
  access to the momentum and the energy dispersion of electronic
  excitations and allows to explore the transition from individual to
  collective excitations. Dimensionality and geometry play thereby a
  key role.  As a prototypical example we analyze theoretically the
  case of Buckminster fullerene C$_{60}$ using \emph{ab initio}
  calculations based on the time-dependent density-functional theory.
  Utilizing the non-negative matrix factorization method, multipole
  contributions to various collective modes are isolated, imaged in
  real space, and their energy and momentum dependencies are traced.
  A possible experiment is suggested to access the multipolar
  excitations selectively via EELS with electron vortex (twisted) beams.
  Furthermore, we construct an accurate analytical model for the
  response function. Both the model and the \emph{ab initio} cross
  sections are in excellent agreement with recent experimental data.
\end{abstract}

\maketitle

%
%
Plasmonics, a highly active field at the intersection of nanophotonics, material
science and nanophysics~\cite{schuller_plasmonics_2010}, has a long
history dating back to the original work of Gustav Mie on light
scattering from spherical colloid particles~\cite{mie_beitrage_1908,halas_plasmons_2011}.
For extended systems the plasmon response occurs at
a frequency set by the carrier density while in a finite system
topology and finite-size quantum effects play a key role.  E.g., for a
nano-shell
\cite{liz-marzan_tailoring_2006,jain_universal_2007,prodan_hybridization_2003}
in addition to the volume mode, two coupled ultraviolet surface
plasmons arise having significant contributions from higher
multipoles, as demonstrated below.  Such excitations can be accessed
by optical means as well as by electron energy-loss spectroscopy
(EELS)~\cite{egerton_electron_2009,garcia_de_abajo_optical_2010}.
 Particle-hole ($p\mhyphen h$) excitations and collective
modes may ``live'' in overlapping momentum-energy
domains and couple in a size-dependent way that cannot be understood
classically~\cite{yannouleas_landau_1992,alasia_single-particle_1994,moskalenko_attosecond_2012}.
%
Giant plasmon resonances were measured in buckminster fullerene
C$_{60}$~\cite{sohmen_electron_1992,burose_electron_1993,
  bolognesi_collective_2012,hertel_giant_1992,scully_photoexcitation_2005,reinkoster_photoionization_2004}
and explained, e.g., by assuming C$_{60}$ to have a constant density
of electrons confined to a shell with inner ($R_1$) and outer ($R_2$)
radii (the spherical shell model)
\cite{ostling_collective_1996,vasvari_collective_1996,verkhovtsev_formalism_2012}.
Refinements in terms of a semi-classical approximation (SCA)
incorporate the quantum-mechanical density extending out of the shell
$R_1<r<R_2$ (so-called spill-out density
\cite{esteban_bridging_2012}). Time-dependent density functional
theory
(TDDFT)~\cite{prodan_structural_2003,marques_time-dependent_2004,esteban_bridging_2012}
was also employed in a number of
calculations~\cite{bauernschmitt_experiment_1998,madjet_photoionization_2008,maurat_surface_2009},
however, most of them use the jellium model, i.e., the ionic structure
is smeared out to a uniform positive background. \\ We present here,
to our knowledge, the first atomistic full-fledge TDDFT calculations
for EELS from C$_{60}$ at \emph{finite} momentum transfer. We
demonstrate the necessity of the full \emph{ab-initio} approach by
unraveling the nature of the various contributing plasmonic modes and
their multipolar character. This is achieved by analyzing and
categorizing the \emph{ab initio} results by means of the
\emph{non-negative matrix factorization} method
\cite{lee_learning_1999}. The results are in line with recent
experimental findings \cite{verkhovtsev_interplay_2012}. The analysis
also allows for constructing an accurate analytical model response
function.

In  first Born approximation for the triply-differential
 cross section (TDCS) for detecting an electron with momentum
$\vec{p}_f$, i.e., measuring its solid scattering angle $\dd\Omega$
and energy $\epsilon_{\vec{p}_f}$ is
\begin{equation}
  \label{eq:crossectiondef}
  \frac{\dd^3\sigma}{\dd\omega\dd\Omega}
  = \frac{4 \gamma^2}{q^4} \frac{p_f}{p_i} S(\vec q,\omega) \ .
\end{equation}
Here,  $\vec{p}_i$ is the incidence momentum corresponding to an energy $\epsilon_{\vec{p}_i}$,
 $\gamma$ is the Lorenz factor,  $\vec q = \vec{p}_f -
\vec{p}_i$ is the momentum transfer, and $\omega =
\epsilon_{\vec{p}_f}-\epsilon_{\vec{p}_i}$
(atomic units are used throughout). $S(\vec q,\omega)$ is the
\emph{dynamical structure factor} akin solely to the
target~\cite{giuliani_quantum_2005}.
\begin{figure}[b]
\includegraphics[width=0.95\columnwidth]{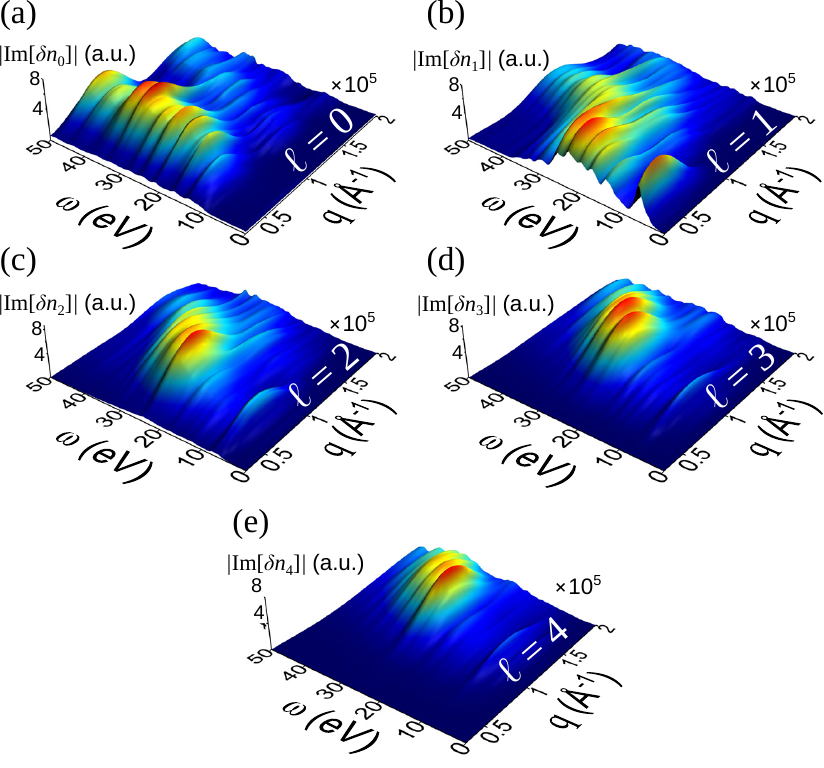}
\caption{(Color online) The $\ell$-resolved constituents of the dynamical structure factor
  of C$_{60}$, $|\mathrm{Im}[\delta n_{\ell}(q,\omega)]|$ for (a)
  $\ell=0$, (b) $\ell=1$, (c) $\ell=2$, (d) $\ell=3$, and
  (e) $\ell=4$.  The C$_{60}$ molecule was treated in standard truncated
  icosahedric geometry with bond lengths $r_\mathrm{C-C}=1.445$~\AA\
  and $r_\mathrm{C=C}=1.390$~\AA. 
  \label{fig1}}
\end{figure}
The fluctuation-dissipation \cite{giuliani_quantum_2005} theorem links
$S(\vec q,\omega)$ with the non-local, retarded density-density linear
response function $\chi^\mathrm{R}(\vec r,\vec r^\prime;t-t^\prime)$
\cite{onida_electronic_2002,giuliani_quantum_2005,marques_fundamentals_2012}
via
$S(\vec q,\omega)=-(1/\pi)\mathrm{Im} [\chi^\mathrm{R}(\vec q,-\vec
q;\omega)]$.
On the other hand, $\chi^\mathrm{R}(\vec r,\vec r^\prime;t-t^\prime)$
describes the change in the system density $\delta n(\vec r ,t)$ upon
a small perturbing potential $\delta \varphi(\vec r,t)$, i.e.
\begin{equation}
  \label{eq:linrespdef}
  \delta n(\vec r,t)=-\int^{\infty}_{-\infty}\!\dd
  t^\prime\!\int\!\dd\vec{r}^\prime \, \chi^\mathrm{R}(\vec r,\vec r^\prime; t-t^\prime)
  \delta \varphi(\mathbf{r}^\prime,t^\prime) \ .
\end{equation}
The response function is determined by evaluating the
density variation with tunable perturbations, as  accomplished
via TDDFT which delivers $\delta n(\vec r,t)$ upon solving the
time-dependent Kohn-Sham (KS) equations \cite{Note1}.

Along this line, we utilized the {\sc Octopus} package
\cite{marques_octopus:_2003,andrade_time-dependent_2012}, and
propagated the KS equations. Kohn-Sham states are represented on a
uniform real space grid \cite{andrade_real-space_2015} (0.2~\AA\ grid
spacing) confined to a sphere with 10~\AA\ radius. For the ground
state we checked the performance of different typical functionals and
found that the local-density approximation (LDA) improved by
self-interaction correction (SIC) yields fairly good results. The HOMO
(-9.2~eV) is located slightly too low with respect to the experimental
value (-7.6~eV) \cite{hertel_giant_1992}. The band width (which is
typically underestimated in DFT) within the LDA+SIC scheme is the
largest for the tested functionals \cite{Note2}. LDA-type
Troullier-Martins pseudopotentials are used to incorporate the
influence of the two core electrons per C atom, such that only the 240
valence electrons accounted for. Gaussian smearing has been employed
to deal with the degeneracy of the HOMO. In gas-phase the molecules
are randomly oriented. Hence, we have to evaluate the spherically
averaged structure factor $S(q,\omega)$. Technically, this can be
accomplished by choosing the perturbation $\delta \varphi(\vec r,t)=
I_0 \delta(t) j_\ell(q r) Y^*_{\ell m}(\Omega_{\vec r})$
\cite{sakko_time-dependent_2010} where $j_\ell$ is the spherical
Bessel function and $Y_{\ell m}$ is the spherical harmonic. The
perturbation strength lies with $I_0=0.01$~a.u. well within the regime
of linear response. The perturbed states are then propagated by the
AETRS propagator \cite{castro_propagators_2004} up to $T=20~\hbar$/eV
with a time step of $\Delta t = 2\times 10^{-3} \hbar$/eV, covering
the range from $0.31$~eV to $3142$~eV in frequency space. The large
simulation box ensured the adequate representation of excited
states. A mask was multiplied to the Kohn-Sham states at each time
step in order smoothly absorb contributions above the ionization
threshold. From the density variation $\delta n(\vec r,t)=n(\vec
r,t)-n(\vec r,t=0)$, $\delta n_{\ell m}(q,t) =\int \dd\vec r\, \delta
n(\vec r,t) j_\ell(q r) Y_{\ell m}(\Omega_{\vec r})$ is then computed
in each time step and Fourier transformed to $\delta n_{\ell
  m}(q,\omega) $ allowing to determine $S(q,\omega)$ as
\begin{equation}
  \label{eq:Ssphav}
  S(q,\omega) = -\frac{4}{I_0} \sum^{\ell_\mathrm{max}}_{\ell=0}\sum^\ell_{m=-\ell}
  \mathrm{Im}\left[ \delta n_{\ell m}(q,\omega) \right] \ .
\end{equation}
The $m$-dependence is subsidiary. 
 To a good approximation henceforth $m=0$ (cf. Eq.~\eqref{eq:Ssphav}).  
 It is sufficient to consider
$|\mathrm{Im}[\delta n_{\ell}(q,\omega)]|\equiv-\mathrm{Im}[\delta
n_{\ell,m=0}(q,\omega)]$ which stands for the $\ell$-resolved
dynamical structure factor depicted in Fig.~\ref{fig1}.  For
$q\rightarrow 0$ (in the optical limit) the dipolar term is clearly
dominant over higher multipoles.\\
According to the shell model \cite{verkhovtsev_formalism_2012} the
C$_{60}$ molecule possesses a volume plasmon mode ($\ell=0$ and radial
density oscillation with one node), a symmetric surface mode ($\ell\ge
1$ and no radial oscillation), and an anti-symmetric surface mode
($\ell\ge 1$ and one radial node). We denote these modes by V, S1 and
S2, respectively. The plasmon energies are derived as
$\omega_{\mathrm{V}} = \sqrt{3/r^3_s}$,
$\omega^2_{\mathrm{S(1,2)},\ell}=\frac{\omega^2_{\mathrm{V}}}{2}\big[1\mp
\frac{1}{2\ell+1}\sqrt{1+4\ell(\ell+1)(R_1/R_2)^{(2\ell+1)}}\big]$.
Inspecting the $\ell=1$ panel the two surface modes may be identified
around $q\sim
0.3$~\AA$^{-1}$, $\omega \sim 20$~eV and $q\sim 1$~\AA$^{-1}$, $\omega\sim 40$~eV. \\
As evident from Fig.~\ref{fig1}, for higher $q$ plasmonic modes
($\mathrm{S1,S2,V}$) seem to merge and attain various multipoles
contributions. This is a manifestation of electronic transitions
between the single-particle states with different angular
momentum~\cite{feng_atomlike_2008,pavlyukh_angular_2009,pavlyukh_communication:_2011}. Thus,
the question arises of how
to  disentangle these modes and to unravel their multipolar nature. \\
\begin{figure}[t]
\includegraphics[width=\columnwidth]{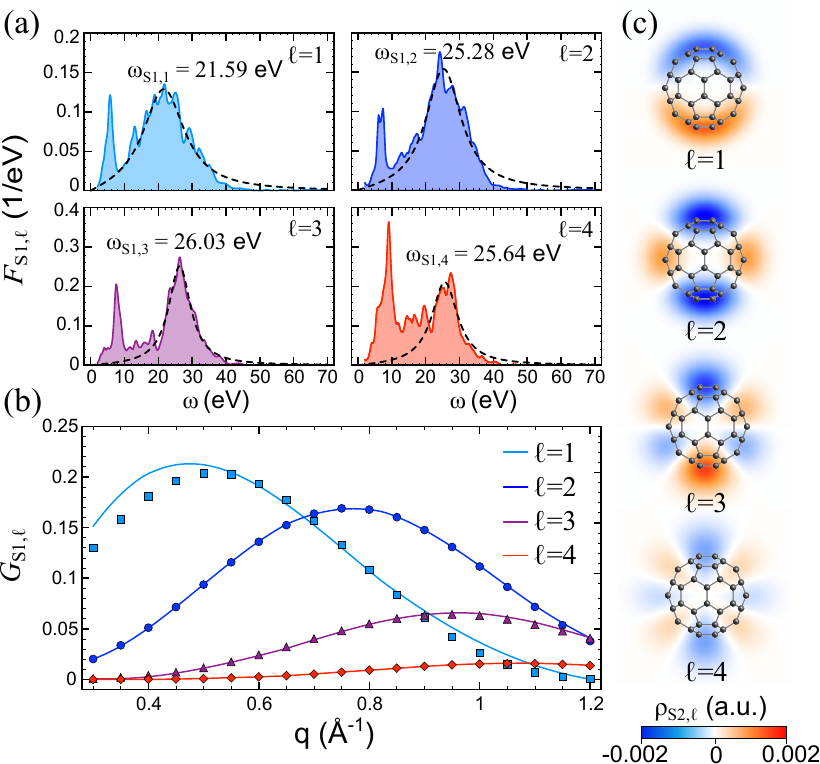}
\caption{(Color online) (a) Frequency-dependent part of the S1 modes
  obtained from the NMF (shaded curves) with fits (dashed lines).  For
  plasmon features we concentrate on the region $\omega>18$~eV. (b)
  $q$-dependent part of the S1 modes from the NMF (solid lines) along
  with fits using the model fluctuation density (symbols).  (c) Model
  fluctuation density $\rho_{\mathrm{S1,\ell}}$ in a plane cut through
  the center of the molecule for $\ell$ ranging from 1 (top) to 4
  (bottom).\label{fig2}}
\end{figure}
%
A suitable mathematical tool to tackle this task is the non-negative
matrix factorization (NMF), which is extensively used, e.~g., for face
recognition algorithms \cite{lee_learning_1999}.  Applied to our
problem, the NMF delivers two functions $F_i(\omega)\ge0$ and
$G_i(q)\ge0$ that enter the density response as $\vert\delta
n_{\ell}(q,\omega)\vert=\sum_i F_i(\omega)G_i(q)$ (see
appendix~\ref{appendixa}). This structure follows namely from the
Lehmann representation of $\chi^\mathrm{R}(\vec r,\vec
r^\prime;\omega)$ as
\begin{equation}
  \label{eq:lehmannrsp}
  \chi^\mathrm{R}(\vec r,\vec r^\prime;\omega) = \sum_\alpha \xi_\alpha(\omega)
  \rho_\alpha(\vec r)\rho_\alpha(\vec r^\prime),\quad \xi_\alpha(\omega)=\frac{2 E_\alpha}{(\omega+\iu
\Gamma_\alpha)^2-E^2_\alpha} \
\end{equation}
where $\rho_\alpha$ is the real fluctuation density corresponding to a
transition from the ground to an excited many-body state labelled by
$\alpha$ (with excitation energy $E_\alpha$), and
 $\Gamma_\alpha$ is the line width. Assuming spherical symmetry,  excitations have
angular ($\ell$) and   radial ($\nu$) components. Expanding
$\rho_\alpha(\vec r)\rho_\alpha(\vec r^\prime)=\sum_{\ell m}
R_{\nu,\ell}(r)R_{\nu,\ell}(r^\prime)Y_{\ell m}(\Omega_{\vec
  r})Y^*_{\ell m}(\Omega_{\vec r^\prime})$ Eq.~\eqref{eq:lehmannrsp}
implies for the structure factor
 \begin{eqnarray}
 && S(q,\omega)=\sum_{\nu\ell} (2\ell+1) F_{\nu,\ell}(\omega)G_{\nu,\ell}(q),\nonumber\\
&&F_{\nu,\ell}(\omega) = \mathrm{Im}[\xi_{\nu,\ell}(\omega)], \quad
G_{\nu,\ell}(q)=\left(\int^\infty_0\!\dd r\, r^2 R_{\nu,\ell}(r) j_\ell(q r)
\right)^2.\nonumber\end{eqnarray}
In full generality the sum \eqref{eq:lehmannrsp} contains infinite
number of terms corresponding to the infinite number of excited
states.  For homogeneous electron gas plasmons are strongly damped
when their momentum enters the $p\mhyphen h$ continuum, where the
non-interacting structure factor $S^{(0)}(q,\omega)>0$. For electrons
confined to a spherical shell the momentum can be represented by a
magnitude $q$ and an angular momentum $\ell$. To mark the effective
region $q_\mathrm{max}$ and $\ell_\mathrm{max}$ in which plasmon modes
exist, we estimate the transverse momentum as $2\ell \pi/R$ (with
radius $R$) and compare it to the critical momentum
$q_\mathrm{crit}=0.559 k_\mathrm{F}$
\cite{stefanucci_nonequilibrium_2013} (the Fermi momentum is
$k_\mathrm{F}=(9\pi/4)^{1/4}r^{-1}_s$).  We find so a critical $\ell
\sim 3$. Thus, any collective excitation beyond $\ell_\mathrm{max}=4$
will be suppressed.  For a complementary picture, we analyzed
$S^{(0)}(q,\omega)$ in SCA \cite{pavlyukh_fast_2012}, for which the
electron density enters as a central ingredient (we take the
spherically-averaged DFT density $n_0(r)$) \cite{Note3}.  This allows
to estimate for which $q$ the $p\mhyphen h$ pairs dominate the
spectrum for each $\ell$ separately.  For $\ell_\mathrm{max}=4$ we
find the $p\mhyphen h$ domain at $q \gtrsim 1.2$~\AA$^{-1}$. Note that
due to geometrical confinement plasmons and $p\mhyphen h$ excitations
intersect each other and couple so significantly. \\
Now we separate the response into $\nu=\text{S1}$ (Fig.~\ref{fig2})
and $\nu=\text{S2}$ (Fig.~\ref{fig3}) for $\ell\ge 1$, while the mode
$\nu=\text{V}$ can be found from $\ell=0$ density component
(Fig.~\ref{fig4}). The onset of $p\mhyphen h$ excitations is also
present in the spectra.
\begin{figure}[t]
  \centering
  \includegraphics[width=\columnwidth]{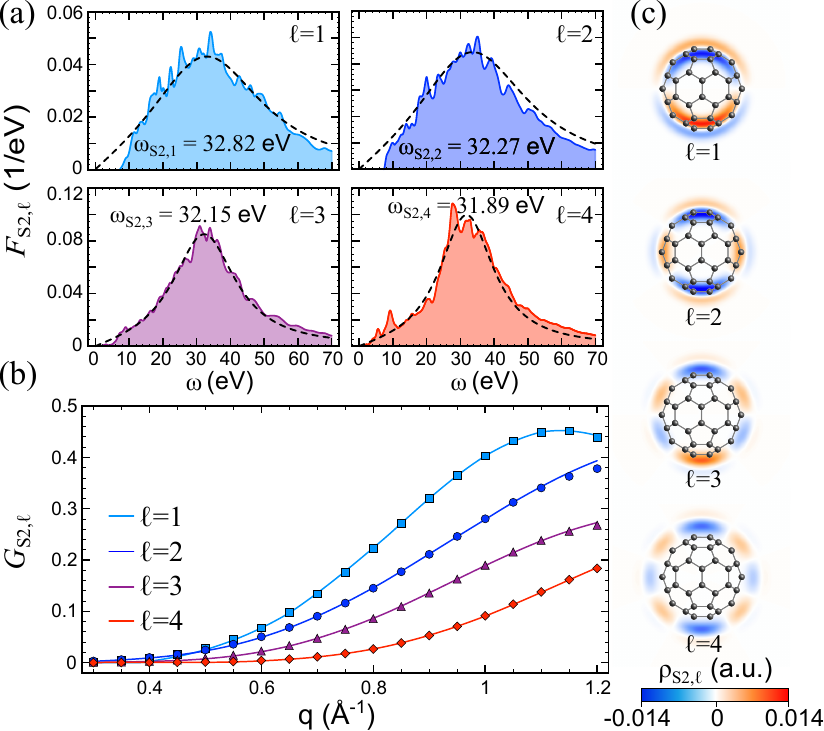}
  \caption{(Color online) (a) Frequency-dependent part of the S2 modes
   from the NMF (shaded curves) and corresponding fits (dashed lines).
   For the latter, no constraint has been imposed on the frequency range.
   (b) $q$-dependent part of the S2
  modes from the NMF (solid lines) and fits (symbols).
  (c) Model fluctuation density $\rho_{\mathrm{S2,\ell}}$ as in Fig.~\ref{fig2}.
  \label{fig3}}
\end{figure}
The plasmon frequencies $\omega_{\nu,\ell}$ are identified from the
maximum of the $\omega$-dependence spectra as obtained by the NMF in
the form
$F^\mathrm{fit}_{\nu,\ell}(\omega)=\mathrm{Im}[2\omega_{\nu,\ell}/((\omega+\iu
\Gamma_{\nu,\ell})^2-\omega^2_{\nu,\ell})]$.  Inspecting
Fig.~\ref{fig2}(a), we find the dipole plasmon at
$\omega_{\mathrm{S1},1}\simeq 21.59$~eV, this is a well established
value. Increasing $\ell$ shifts the peak to larger energies (in line
with the shell model); the sharp peak around 7.5~eV, which is known to
consist of a series of $p\mhyphen h$ excitations
\cite{moskalenko_attosecond_2012}, gains spectral weight until it
dominates for $\ell=4$. Abundance of large angular momentum states
around HOMO-LUMO gap \cite{pavlyukh_communication:_2011} increases
the number of channels for high-multipole electronic transitions and
is responsible for the peak's enhancement. The plasmon frequency
$\omega_{\mathrm{S1},4}=25.64$~eV on the other hand is smaller than
$\omega_{\mathrm{S1},3}=26.03$~eV. This demonstrates the limitations
of the SCA.

The radial profile of the density oscillations $R_{\nu,\ell}(r)$ can be inferred from
$G_{\nu,\ell}(q)$ in that we assume
 $R^{\mathrm{fit}}_{\mathrm{S1},\ell}(r)= A_\ell r \exp[-(r-r_\ell)^2/2\sigma^2_\ell]$
and extract the parameters $(A_\ell,r_\ell,\sigma_\ell)$ for which $\big\| G^\mathrm{}_{
  \mathrm{S1},\ell}(q)-\left(\int^\infty_0\!\dd r\, r^2
R^\mathrm{fit}_{\mathrm{S1},\ell}(r) j_\ell(q r) \right)^2 \big\|$ is minimized.
The effective fluctuation densities are then given by
$\rho_{\mathrm{S1},\ell}(\vec r) = R_{\mathrm{S1},\ell}(r) Y_{\ell
  0}(\Omega_{\vec r})$, cf.  figure~\ref{fig2}(c).

\begin{figure}[t]
  \centering
  \includegraphics[width=\columnwidth]{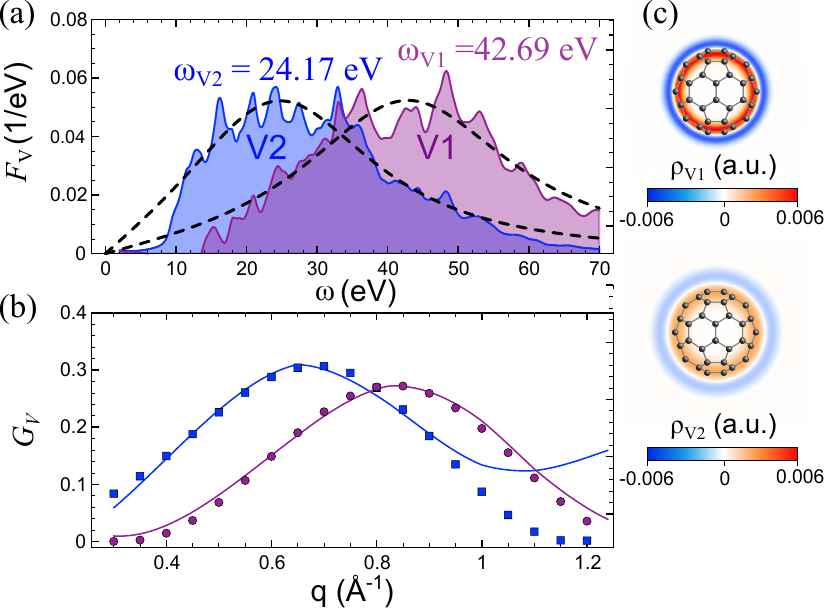}
  \caption{(Color online) (a) Frequency-dependent part of the V1 (blue shaded curve) and
    V2 (purple shaded curve) mode from the NMF along with corresponding fits (dashed
    lines). Fitting has been carried out in the complete frequency range.  (b)
    $q$-dependent part of the V1 and V2 modes from the NMF (solid lines) and fits
    (symbols). (c) Model fluctuation densities in the same plane as in
    Fig.~\ref{fig2}. \label{fig4}}
\end{figure}

Analogous procedure for S2 modes (Fig.~\ref{fig3}) reveals a decrease of the plasmon
energies in qualitative agreement with Ref.~\cite{bolognesi_collective_2012}. However,
the dispersion is less pronounced than in the shell model.
To characterize the
fluctuation densities, we use an Ansatz containing a node $R^{\mathrm{fit}}_{
  \mathrm{S2},\ell}(r)= A_\ell r (1-r/r^{(0)}_\ell)\exp[-(r-r_\ell)^2/2\sigma^2_\ell]$ and
determine the parameters as to match $G_{\mathrm{S2},\ell}(q)$ (Fig.~\ref{fig3}(b)). The
spatial structure of the plasmon oscillation is shown in Fig.~\ref{fig3}(c).

A common and physically intuitive feature of the S1 and S2 modes is
that the spatial extend of the fluctuation density is growing
with $\ell$. This is a consequence of the increasing centrifugal
force, "pushing" the oscillation away from the center.

Applying the NMF with two components to $|\mathrm{Im}[\delta
n_{0}(q,\omega)]|$ shows (Fig.~\ref{fig4}) that in addition to the
expected volume plasmon (labelled by V1) around
$\omega_{\mathrm{V1}}=42.69$~eV (which agrees well with density
parameter $r_s \sim 1$), a second resonance peaked around
$\omega_{\mathrm{V2}}=24.17$~eV appears. To clarify its origin we
computed the response function from its non-interacting counterpart in
the random-phase approximation and invoking the SCA (see
appendix~\ref{appendixb}). After obtaining $|\mathrm{Im}[\delta
n_{0}(q,\omega)]|$ we applied the NMF, as well. This procedure yields
very similar spectra including the occurrence of V2. This feature is,
however, very sensitive to the details of the density distribution; it
vanishes for a discontinuous step-like profile.  Thus, it is the
oscillations of the spill-out density taking place on the surface of
the molecule that form V2. This
is a pure quantum effect. \\
\begin{figure}
  \centering
  \includegraphics[width=\columnwidth]{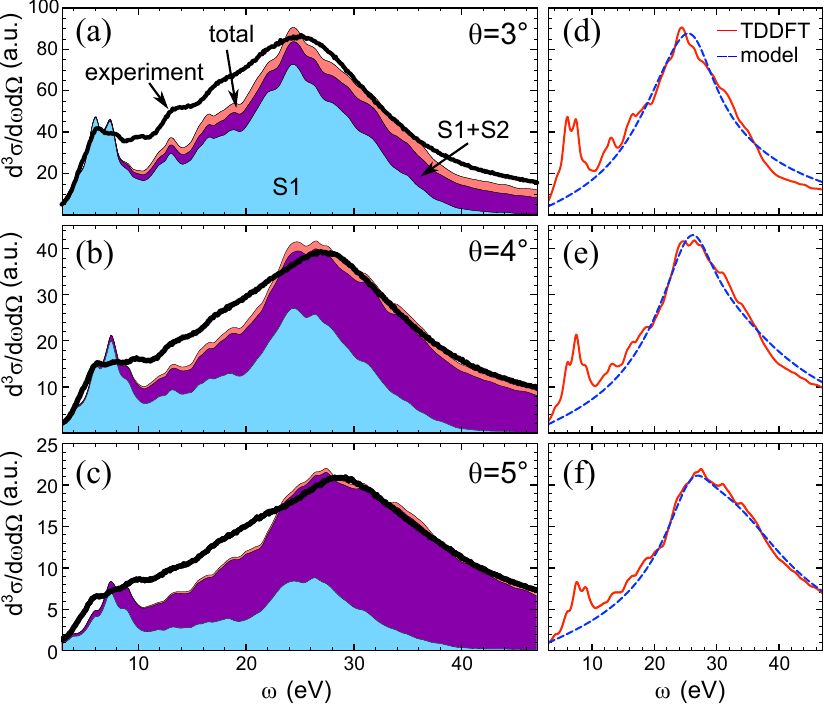}
  \caption{(Color online) TDCS for EELS of C$_{60}$ at scattering
    angles $\theta=3^\circ$ (a), $\theta=4^\circ$ (b), and
    $\theta=5^\circ$ (c). The energy loss $\omega$ is with respect to initial beam
    energy of $\epsilon _0=1050$~eV. Colored curves represent TDDFT
    calculations resolved in the contribution from S1, S1+S2 and
    S1+S2+V. The thick curve shows experimental data \cite{verkhovtsev_interplay_2012}.
    (d--f): comparison of full TDDFT and model cross sections.
    \label{fig5}}
\end{figure}

With the dynamical structure factor being fully characterized, we
proceed by computing the TDCS (Eq.~\eqref{eq:crossectiondef}).
Fig.~\ref{fig5} compares calculated and measured
\cite{verkhovtsev_interplay_2012} EELS spectra as a function of the
electron scattering angle $\theta$ which fixes the momentum
transfer. The magnitudes of the measured spectra shown in
Fig.~\ref{fig5} are determined up to an overfall factor.  Thus, the
theory-experiment comparison in Figs.~\ref{fig5} (b,c) is on an
absolute scale.  The classification of the plasmon modes accomplished
by the NMF analysis allows for plotting mode-resolved TDCS curves. As
Figs.~\ref{fig5}(a--c) demonstrate, the S1 plasmons play the dominant
role for small $\theta$ (which corresponds to the optical limit of
small $q$), while the S2 modes becomes increasingly significant for
larger $\theta$ (i.e., larger $q$). The larger energy of the S2 with
respect to the S1 plasmons leads to the formation of a shoulder
(clearly visible for $\theta=4^\circ$) and, thus, to the apparent
shift of the maximum of the experimental EELS spectrum with growing
$\theta$. A similar effect is also observed for the S1 modes due to
their dispersion with
respect to $\ell$.\\
 Furthermore, the extracted $\omega$-dependencies and the model
 fluctuation densities can be used to construct an  approximate
 structure factor $S^\mathrm{model}(q,\omega)=\sum_{\nu\ell}
 (2\ell+1) F^\mathrm{fit}_{\nu,\ell}(\omega)
 G^\mathrm{fit}_{\nu,\ell}(q)$ that reproduces the TDDFT results
 around the plasmon resonances in a precise way by
 construction. Corresponding TDCSs are compared in of
 Figs.~\ref{fig5}(d--f).\\
 An important feature of the structure factor is the $f$-sum rule
 $\int^\infty_0\!\dd\omega \, \omega S(q,\omega) = N q^2/2$ (number of
 electrons $N$). Checking for the (plasmon-dominated)
 $S^\mathrm{model}(q,\omega)$ shows the discrepancy for larger $q$; a
 critical value is reached when $\int^\infty_0\!\dd\omega \, \omega
 S^\mathrm{model}(q,\omega)$ decreases again after quadratic
 growth. We find $q_\mathrm{crit}\sim 1.2$~\AA$^{-1}$ which is
 consistent with the estimation above. Hence, $p\mhyphen h$
 excitations become more important for $q > q_\mathrm{crit}$ and
 gradually diminish the plasmon contribution.

 In summary, we presented accurate TDDFT calculations for the
 dynamical structure factor and EELS spectra for C$_{60}$ molecule
 underlining the role of higher multipole contributions. Using NMF
 decomposition allowed to trace the evolution in $q$ and $\omega$ of
 the symmetric and anti-symmetric surface and volume plasmons. In
 addition, we characterized and modeled the fluctuation densities
 (i.e., the ingredients of the response function) and unveiled their
 multipolar character.  These ingredients might, in principle, be
 accessed selectively by using electron beams carrying a definite
 angular momentum (electron vortex beams
 \cite{uchida_generation_2010,verbeeck_production_2010}). By measuring
 the angular momentum of the scattered beam the \emph{angular momentum
   transfer} $\Delta \ell$ becomes a control variable which the EELS
 spectra depends on. Particularly, provided the beam axis coincides
 with the symmetry axis of spherical system, the plasmonic response
 upon scattering of such twisted electrons contains multipole
 contributions for $\ell \ge |\Delta \ell|$ only \cite{Note4}.  Hence,
 specific multipoles can be excluded or included by varying $\Delta
 \ell$.

 Furthermore, we discussed the limitation of spherical-shell models in
 describing the quenching of the volume plasmon and identified the
 electronic density distribution as a key factor determining its
 energy. We obtained excellent agreement with experimental results and
 explained how the different plasmon modes contribute to the spectra.


\appendix

\section{Non-negative matrix factorization\label{appendixa}}

As dictated by the fluctuation-dissipation theorem, the imaginary part
of $\delta n_{\ell} (q,\omega)$ for $\omega >0$ is purely
negative. Thus, the non-negative matrix factorization (NMF) can be
applied to $|\mathrm{Im}[\delta n_{\ell}(q,\omega)]| =
-\mathrm{Im}[\delta n_{\ell}(q,\omega)]$ to split
\begin{equation}
  \label{eq:nmf}
  \left|\mathrm{Im}\left[\delta n_{\ell}(q,\omega)\right]\right|
  = \sum^{N}_{\nu=1} F_{\nu,\ell} (\omega) G_{\nu,\ell}(q) \ .
\end{equation}
Without imposing any restriction on the number of components ($N$)
the expansion \eqref{eq:nmf} is exact and can be paralleled with the
singular value decomposition (SVD) of a general (complex or real)
matrix $\mathbf{M}$:
$\mathbf{M}=\mathbf{U}\mathbf{\Sigma}\mathbf{V}^*$. The difference is
in the additional requirements of positivity on the vectors forming
$\mathbf{U}$ and $\mathbf{V}$. The transition from continuous
variables as in eq.~\eqref{eq:nmf} to the matrix form is provided by
discretizing the $\omega$- and the $q$-points after smooth
interpolation.

We select $N=2$ as we expect two dominant surface plasmon modes (S1
and S2). This choice is confirmed by computing the residue norm with
respect to the full function $|\mathrm{Im}[\delta
n_{\ell}(q,\omega)]|$.

The problem of non-negative matrix factorization can be formulated as
a \emph{non-convex} minimization problem for the residue norm
$r=||\vec{A}-\vec{W}\vec{H}||^2$. Thus, the solution is not unique and
may lead to \emph{local minima}. Depending on the norm used different
algorithms can be formulated. A commonly used method is the
\emph{multiplicative update} of D.~Lee and
S.~Seung~\cite{lee_learning_1999}:
\begin{subequations}
\label{eq:mu_rules}
\begin{eqnarray}
\vec{W}_{ia}&\leftarrow&\vec{W}_{ia}\frac{(\vec{A}\vec{H}^T)_{ia}}{(\vec{W}\vec{H}\vec{H}^T)_{ia}},\\
\vec{H}_{aj}&\leftarrow&\vec{H}_{aj}\frac{(\vec{W}^T\vec{A})_{aj}}{(\vec{W}^T\vec{W}\vec{H})_{aj}},
\end{eqnarray}
\end{subequations}
where $i$ indexes the energy points and $j$ numbers the time points. The method starts
with some suitable guess for matrices $\vec W$ and $\vec H$. Additionally, the vectors
forming $\vec W$ are normalized each step:
\[
\vec{W}_{ia}\leftarrow \frac{\vec{W}_{ia}}{\Vert\vec{W}_a\Vert}.
\]
Upon these prescriptions~\eqref{eq:mu_rules} the Euclidean distance
$r$ monotonously decreases until the stationary point (local minimum)
has been reached. We initialized the vector $\vec W_{1}$ ($\vec W_2$)
with cuts of $|\mathrm{Im}[\delta n_{\ell m}(q,\omega)]|$ along $q$
direction at $\omega = 20$~eV ($\omega = 40$~eV), while $\vec H_1$
($\vec H_2$) is constructed by cuts at $q=0.5$~\AA$^{-1}$
($q=1.0$~\AA$^{-1}$). We found that typically 1000 iterations yield
well converged results.

The functions $F_{\nu,\ell} (\omega)$ and $G_{\nu,\ell}(q)$ is then
obtained from interpolating the data from $\vec H_\nu$ and $\vec
W_\nu$, respectively. We normalize the frequency spectra such that
fitting by $F^\mathrm{fit}_{\nu,\ell}(\omega) =
\mathrm{Im}[2\omega_{\nu,\ell}/((\omega+\iu
\Gamma_{\nu,\ell})^2-\omega^2_{\nu,\ell})]$ (as explained in the main
text) can be performed without any additional
prefactor. $G_{\nu,\ell}(q)$ is normalized accordingly. This
normalization procedure is consistent with the Lehmann representation.

\section{Semi-classical calculations \label{appendixb}}

In order to eludicate the behavior of the volume plasmons,
semi-classical calculations provide some insight. The starting point
is the Dyson equation for the density-density response function in
random-phase approximation (RPA):
\begin{equation}
  \label{eq:RPA}
  \begin{split}
    \chi(\vec r,\vec r^\prime;z) = \chi^{(0)}(\vec r,\vec r^\prime;z)
    + &\int\!\dd \vec{r}_1 \int\!\dd \vec{r}_2\,  \chi^{(0)}(\vec r,\vec r_1;z) \\
    &\times   v(\vec r_1-\vec r_2) \chi(\vec r_2,\vec r^\prime;z) \ .
  \end{split}
\end{equation}
We drop the superscript R and consider general complex argument $z$
here. In SCA, the non-interacting reference response function
$\chi^{(0)}(\vec r,\vec r^\prime;z)$ can be expressed in terms of
ground-state density $n_0(\vec r)$ ($=n_0(r)$ as we assume spherical
symmetry here) only. The subsequent derivations and the solution
scheme for eq.~\eqref{eq:RPA} are detailed in the Supplementary
Material \cite{Note5}. The amount of spill-out density can be adjusted
by varying the smearing parameter $\Delta r$ in the model density
\begin{equation}
  \label{eq3}
  \begin{split}
    n_0(r) &= N_0 \big[\theta_{\Delta r}(r-R_1) - \theta_{\Delta r}(r-R_2)\big] \ ,\\
    \theta_{\Delta r}(r) &= \frac{1}{1+\exp[-r/\Delta r]} \ ,
  \end{split}
\end{equation}
where $R_1 = R_0 -\Delta R/2$, $R_2 = R_0+\Delta R/2$ are the inner
and outer radii ($R_0 = 6.5$~a.u.), while the normalization $N_0$
ensures the correct total valence charge. $\Delta R$ is fixed to keep
the mean density constant. The scenario $\Delta r \rightarrow 0$
corresponds to a box-like density profile with sharp boundaries, while
$\Delta r = 0.5$~a.u. is a good approximation to the
spherically-averaged DFT density. Once eq.~\eqref{eq:RPA} is solved
for certain $\Delta r$, the $(\ell=0)$ contribution to the structure
factor, $\mathrm{Im}[\delta n_0(q,z=\omega+\iu \Gamma)]$ ($\Gamma =
0.1$~a.u. is a broading parameter) can be computed. Applying the NMF
technique allows again for separating the V1 and V2 modes. We find the
position of V1 similar to the TDDFT results, while the behavior of V2
is very sensititve to $\Delta r$. While very pronounced for $\Delta r
= 0.5$~a.u., the relative strength of the V2 peak vanishes for $\Delta
r\rightarrow 0$. More details and graphs of volume plasmon spectra can
be found in the Supplementary Material.


\begin{acknowledgments}
This work is supported by the DFG under Grants No.  SFB762 and No. PA
1698/1-1. We thank Paolo Bolognesi and Lorenzo Avaldi for fruitful
discussions and for providing experimental data.
\end{acknowledgments}

\end{document}